\documentclass{aa}  

\usepackage{graphicx}
\usepackage{txfonts}
\usepackage[flushleft]{threeparttable}
\begin{document}

   \title{ExoplANETS-A: a VO database for host stars and planetary systems }
   \subtitle{The effect of XUV on planet atmospheres}
\titlerunning{Exoplanets-A Database. XUV effect on planet atmospheres}
   \author{M.~Morales-Calder\'on
          \inst{1}
          \and
          S.R.G.~Joyce
          \inst{2}
          \and
          J.P.~Pye
          \inst{2}
          \and
          D.~Barrado
          \inst{1}
           \and
          M.~Garc\'ia Castro
          \inst{1}
          \and
          C.~Rodrigo
          \inst{1}
        \and
        E.~Solano
        \inst{1}
        \and
        J.D.~Nichols
        \inst{2}
              \and
	    P.O.~Lagage
	    \inst{3}
     \and
          A. Castro-Gonz\'alez\inst{1}
     \and
          R. A. Garc\'{\i}a\inst{3}
     \and
          M. Guedel\inst{4}
    \and
        N. Huélamo\inst{1}
    \and
        Y. Metodieva\inst{4}
    \and
        R.  Waters\inst{5,  6}
          }

   \institute{Centro de Astrobiolog\'{\i}a      (CSIC-INTA), ESAC Campus, E-28692 Villanueva de la Cañada, Madrid, Spain
              \email{mariamc@cab.inta-csic.es}
         \and
         	School of Physics \& Astronomy, University of Leicester, University Road, Leicester, LE1 7RH, United Kingdom 
         \and
             Universit\'e Paris-Saclay, Universit\'e Paris cit\'e, CEA, CNRS, AIM, 91191, Gif-sur-Yvette cedex, France
        \and
            Department of Astrophysics, University of Vienna, T\"urkenschanzstr. 17, 1180 Vienna, Austria     
        \and
           Department of Astrophysics/IMAPP, Radboud University, PO Box 9010, 6500 GL Nijmegen,The Netherlands
        \and
            SRON Netherlands Institute for Space Research, Niels Bohrweg 4, NL-2333 CA Leiden, the Netherlands
\\
            }

   \date{Received September 15, 1996; accepted March 16, 1997}

 
  \abstract
   {ExoplANETS-A is an EU Horizon-2020 project with the primary objective of establishing new knowledge on exoplanet  atmospheres. Intimately related to this  topic is the study of the host-stars radiative properties in order to understand the environment in which  exoplanets lie.}
   {The aim of this work is to exploit archived data from space-based observatories and other public sources to produce uniform sets of stellar data that can establish new insight on the influence of the host star on the planetary atmosphere. We have compiled X-ray and UV luminosities, which affect the formation and the atmospheric properties of the planets, and stellar parameters, which impact the retrieval process of the planetary-atmosphere's properties and its errors.}
   {Our sample is formed of all transiting-exoplanet systems observed by HST or Spitzer. It includes 205 exoplanets and their 114 host-stars. We have built a catalogue with information extracted from public, online archives augmented by quantities derived by the Exoplanets-A work. With this catalogue we have implemented an online database which also includes X-ray and OHP spectra and TESS light curves. In addition,  we have developed a tool, exoVOSA, which is able to fit the spectral energy distribution of exoplanets.}
   {We give an example of using the database to study the effects of the host-star high-energy  emission on the exoplanet atmosphere. The sample has a planet radius valley which is located at 1.8~$\mathrm{R_{\oplus}}$, in agreement with previous studies. Multiplanet systems in our sample were used to test the photoevaporation model and we find that out of 14 systems, only one significant case poses a contradiction to it (K2-3). In this case, the inner planet of the system is above the radius gap while the two exterior planets are both below it. This indicates that some factor not included in the photoevaporation model has increased the mass-loss timescale of the inner planet. In summary, the exoplanet and stellar resources compiled and generated by ExoplANETS-A form a sound basis for current JWST observations and for future work in the era of Ariel.}
   {}


\maketitle
%

\section{Introduction} \label{sec_intro}

With the James Webb Space Telescope, it is possible to observe the composition and structure of exoplanet atmospheres using transit spectroscopy. In order to successfully model the exoplanet atmosphere, it is necessary to have a sound knowledge of the host star, and particularly its radiation output. To this end, we have built a coherent and uniform database of the relevant properties of host stars from online archives (e.g. XMM-Newton XSA, Gaia DR2, SIMBAD) and publications which are cited in the relevant sections. These exoplanet and host-star catalogues are accompanied by computer models to assess the importance of star--planet interactions, for example the `space weather' effects of the star on its planetary system \citep{Strugarek_2022}. The knowledge gained from this project is being published through peer-reviewed scientific journals, and modelling tools have been publicly released\footnote{Full publications list at https://www.explore-exoplanets.eu/resource/project-publications/}.

Seven institutes\footnote{CEA Saclay, Paris, France; CAB-INTA, Madrid, Spain; MPIA, Heidelberg, Germany; University College London, U.K.; University of Leicester, U.K.; SRON, Utrecht, NL; Universitat Wien, Austria} 
in Europe combined their expertise in the field of exoplanetary research to develop the European Horizon-2020 
ExoplANETS-A\footnote{{\it Exoplanet Atmosphere New Emission Transmission Spectra Analysis}; 
https://explore-exoplanets.eu/ ;The ExoplANETS-A project has received funding from the EU's Horizon-2020 programme; Grant Agreement no.~776403, during the time period 2018 -- 2021.} 
project \citep{Lahuis20} under the coordination of CEA Saclay. In the framework of the project, new data calibration and spectral extraction tools, as well as novel parameter retrieval tools, based on 3D models of exoplanet atmospheres, have been developed to exploit archival data from space- and ground-based observatories, and have produced a homogeneous and reliable characterization of the atmospheres of transiting exoplanets.The project has six work packages (WPs); the focus in this paper is on the WP `Host-star properties: the active environment of exoplanets', and more specifically the 
database containing the compiled catalogue, associated data products, and the web-based user interface, which can be found at \textit{http://svo2.cab.inta-csic.es/vocats/exostars/}.

Properties of the exoplanet host stars are key factors to understanding the environment within which the exoplanet lies and estimating the energy inputs to its atmosphere. This influence comes by two different ways: the basic properties of the star have an obvious effect on the planet and in particular, X-ray and UV luminosity affect the formation and the atmospheric properties of the planets \citep{Johnstone_2020, Becker_2020, Poppenhaeger_2021}; on the other hand, stellar parameters such as its luminosity, effective temperature and radius have an impact in the retrieval process of the planetary-atmosphere properties and on its errors \citep{Andersen_2015}. Therefore, a fully characterized host star is a mandatory first step for ExoplANETS-A and for its main goal.

The paper is structured as follows.
Section \ref{sec_cat_cons} describes the compilation of the catalogue from the various data sources. Section \ref{sec_database} describes the database and its associated software tools, validation of the contents, and examples of statistics drawn from the database. Section \ref{sec_discussion} discusses some uses of the database in studying the influence of the host star on the exoplanet, while Section \ref{sec_conclusions} summarises the main conclusions. Appendix \ref{app_dbexamples} gives an example of accessing the database via the web-based user interface.

\section{Compilation \& construction of the host-stars catalogue} \label{sec_cat_cons}

\subsection{Framework}

\begin{figure}
   \centering
   \includegraphics[width=\hsize]{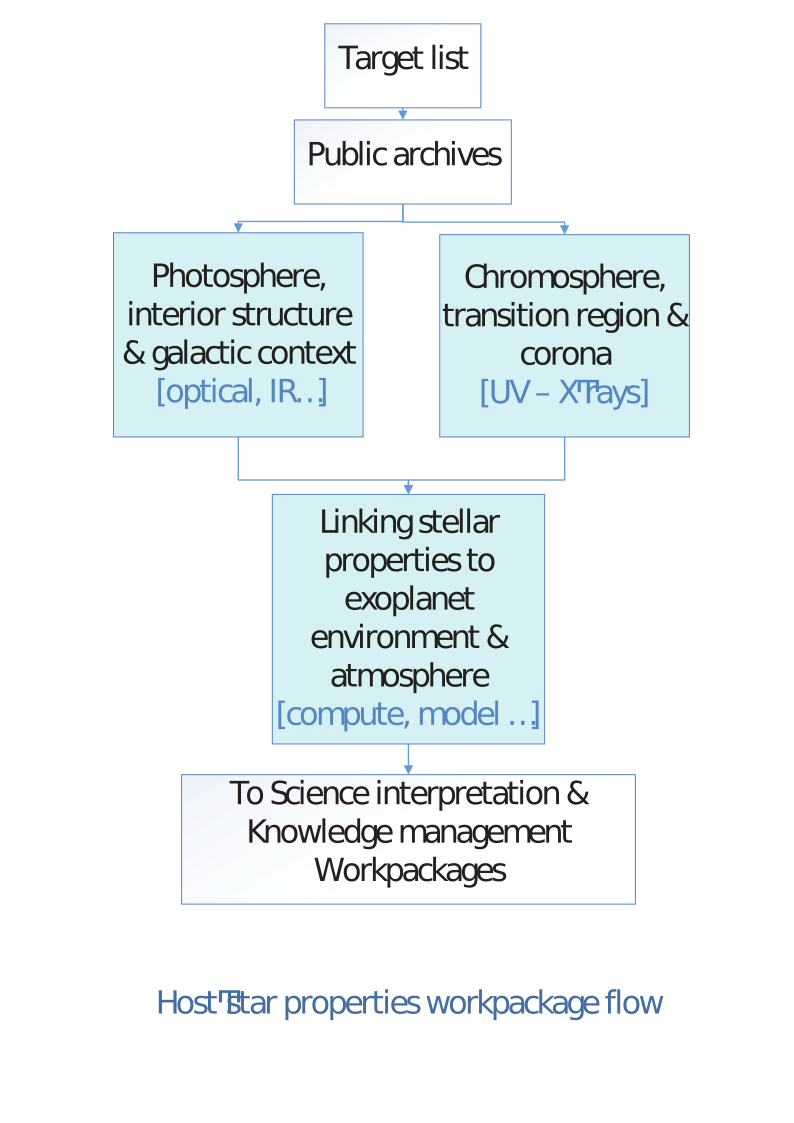}
      \caption{Work and data-flow for the work package 'Host-star properties: The active environment of exoplanets' in the ExoplANETS-A project. 
             }
         \label{fig_dataflow}
   \end{figure}

\begin{figure}
\includegraphics[width=88mm]{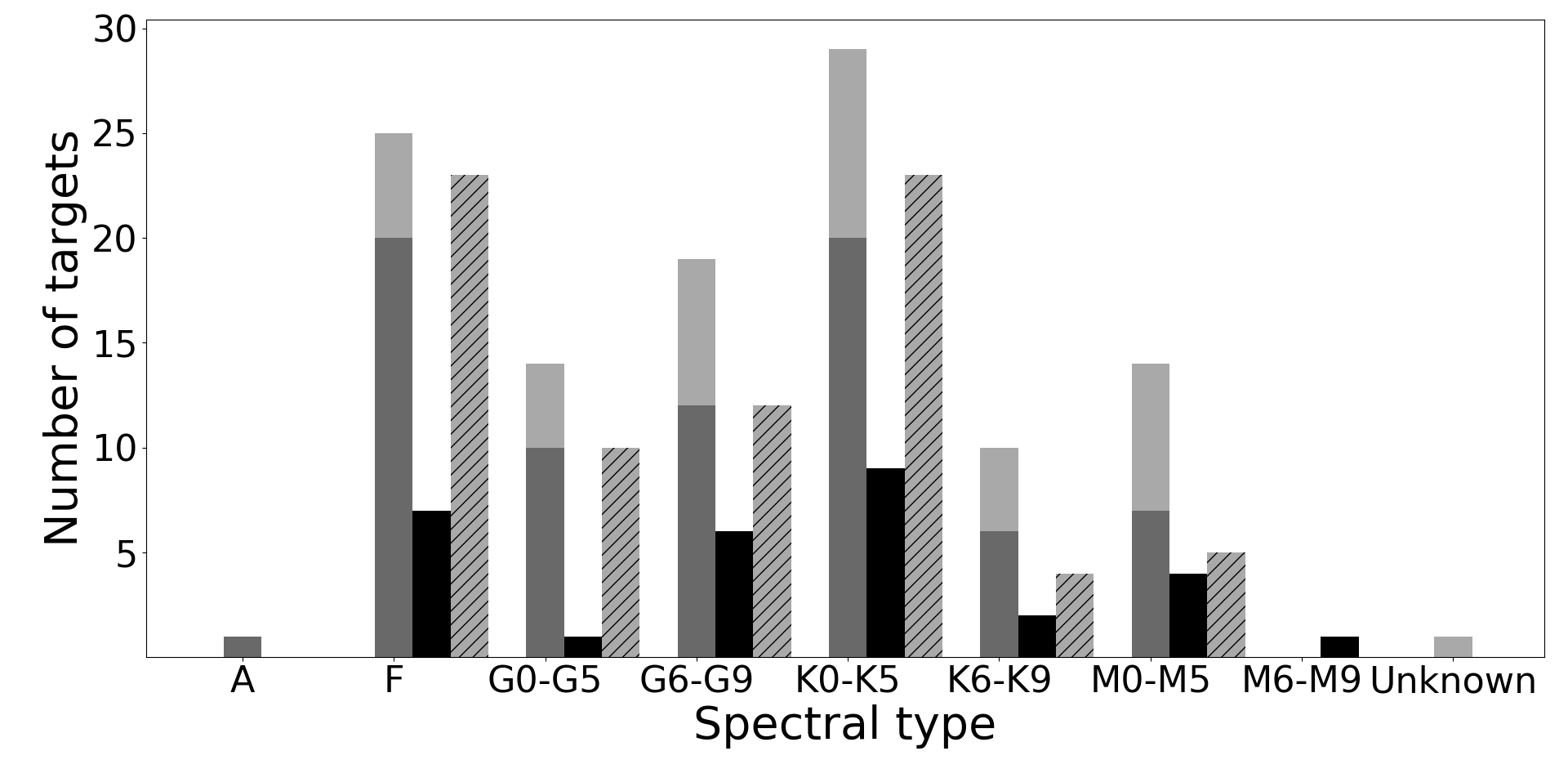}
  \caption{Number of stars in the database which have been detected in X-ray and UV observations and have flux measurements. The histogram shows the distribution of XUV detections according to stellar spectral type. For each spectral type there are 3 bars. The left bar shows the number of targets with HST data (dark grey) and those with only Spitzer data (light grey). The middle bar is the number of X-ray detected targets (black). The right bar is targets with an NUV detection (striped). The left-hand bar can be used to find the total number of targets in a given spectral type. The star with unknown spectral type is \textit{Kepler-14} which is listed as an eclipsing binary in SIMBAD and only has a Gaia photometry value available for one of the stars.}
  \label{fig:XUV_det_histogram}
\end{figure}

The stellar-properties catalogue comprises a compilation of information extracted from public, online astronomical archives (databases and publications), augmented by quantities derived from them by the ExoplANETS-A work. The overall work- and data-flow are shown in schematic form in Fig.\ref{fig_dataflow}.
The database and associated website\footnote{http://svo2.cab.inta-csic.es/vocats/exostars/} were publicly released in December 2019, based on, and updated with information available from public, online sources up to $\sim$October 2020, for the ExoplANETS-A target list. 
In addition, the database contains quantities derived from the ExoplANETS-A project’s own (re)analysis of some of the public data.
The sample of exoplanets and host stars considered by the project comprises all transiting-exoplanet systems observed by Hubble or Spitzer space telescopes until June 2019. This corresponds to 205 exoplanets, of which 121 have HST data; the associated number of stars is 114, with 77 having HST data (Fig. \ref{fig:XUV_det_histogram}). 

\subsection{Optical/IR data}
Precise knowledge of stellar parameters is needed to fully characterize the planet properties. For example, precise stellar  radii are critical if we want  to  measure  precise  values  for  the  radius  of  a  transiting planet. The determination of stellar  radii  depends  in  turn  on  the  quality  of  the  derived  stellar parameters such as the effective temperature. The chemical composition of a planet  is also related to the chemical composition of the protostellar cloud, and thus, of the stellar atmosphere. Thus the characterization of the host stars should be as accurate as possible. Since there are no accurate values for all the parameters, we have tried to, at least, compile uniform sets of stellar parameters to avoid many different teams making use of different methods to derive the same parameters. This is also true for the X-ray and UV data in the next subsection. In this way, the main sources for our catalogue have been SWEET-Cat (atmospheric parameters, \citet{Santos13}), Gaia (astrometric properties, \citet{Gaia}) and SIMBAD (spectral types etc). SWEET-Cat  is a catalogue of uniform stellar atmospheric parameters and masses determined from high-resolution and high S/N spectra for stars with planets. Only one of our planets did not have information in that catalogue, \textit{Kepler}-1625. For the positions and distances we used Gaia DR2 \citep{GaiaDR2} which included all our stars except \textit{Kepler}-14. We collected all the available fluxes at different bandpasses available through Virtual Observatory (VO) services to be able to build the spectral energy distributions for each target and, finally, we completed the dataset with SIMBAD.   

\subsection{X-ray and UV data}
X-ray and UV radiation are produced by the host-star corona and chromosphere and are detected by several space-based observatories.

For many targets, no significant X-ray detection is available. An X-ray upper limit can be estimated from the ROSAT all sky survey. These were gathered from the ESA Upper limit server \citep{Saxton_2011} when available, yielding X-ray flux upper limits for 107 host stars. The Upper Limit server returns an estimated flux when a source is detected, and an upper limit when the source was observed but not detected. It searches both the ROSAT all sky survey \citep{Voges_1999} and the XMM slew survey \citep{Saxton_2008}, as well as pointed observations.

The X-ray data are primarily from XMM-Newton pointed observations, along with several older Einstein and Exosat detections. The existing XMM observations were gathered from the XMM Science Archive, and observations of some new targets (Kepler 138, K2-3, HD3167) were obtained as part of observing program 084441 and 086206 (PI Joyce). Thirty targets have X-ray flux measurements available. The number of targets detected by each satellite (both in X-rays and UV) are listed in Table 1. Some targets are detected by more than one satellite so the total number of detections is more than the 30 unique targets, 26 in the 0.2 - 12 keV energy range (1 - 62 \AA), and 4 Exosat measurements in the 2 - 12 keV range. 
The majority of coronal emission is in the 0.2-2 keV range so the Exosat measurements, which only cover the 2-12 keV range, should be treated with caution. The Upper Limit server includes the Exosat low-energy survey (0.2-2keV), and these are included in the database where available. However, for the 4 targets mentioned, only 2-12 keV survey data were available. 

The XMM-Newton X-ray flux values are recommended when available, due to the high sensitivity of the observatory and the relatively large and uniform set of available results. Detailed spectral fitting with 1- or 2-temperature APEC \citep{Smith_2001} optically-thin plasma models was used to derive the X-ray flux from XMM-Newton spectra. For the values gathered from the Upper Limit server, a standard blackbody model with appropriate temperature (0.3 keV) was used for the flux estimate because an optically-thin plasma model is not implemented in the Upper Limit server. Fitting XMM spectra in \textsc{xspec} indicated that when using a blackbody model, the 0.3 keV temperature yielded a similar count-rate/flux ratio as the APEC model. We found that for the Exosat detections in the 2-12 keV range that a 0.3 keV blackbody resulted in an unrealistically high flux estimate, so the Exosat flux values in the catalogue are based on a 0.1 keV blackbody.

The UV data were gathered from the Galex all-sky surveys \citep{Bianchi_2017}, which provide a flux and magnitude in the NUV (1770 - 2730 \AA) range for 69 out of 114 stars. 23 stars also have an FUV (1350 - 1780 \AA) flux available. The Galex data consist of broad-band photometry. 
Further UV data were gathered from the Swift (Neil Gehrels Swift Observatory,  \citealt{Gehrels_2004}) and XMM-Newton optical monitor \citep{Mason_2001}. 
 Another 7 targets were observed by the XMM-OM but were either saturated or not detected. Details where this occurred for each target are recorded in the database. The Swift UVOT and XMM-OM data consist of photometry in several bands depending on the filter(s) used for a particular observation. The wavelength range relevant to each Swift or XMM-OM observation is given in the database (e.g. the column named: "column inst range Swift UVOT"). Galex data are recommended for studying the sample as a whole because data are available for most targets and provide a uniform analysis. Swift UVOT and XMM-OM UV data are useful as they are usually more recent than the Galex surveys which ended in 2012. Also, UV data are simultaneous with the X-ray data in the case of XMM-Newton.


\begin{table}
\begin{minipage}{88mm}

\caption{Observatories used to gather XUV data for stars in the catalogue.}
\label{table:observations_2018}

\begin{threeparttable}
\begin{tabular}{ccc}
\hline 
\textbf{Observatory} & \textbf{Wavelength ranges} & \textbf{Number of } \\ 
 & (\AA\ or keV) & \textbf{detected targets} \\
\hline 
\textbf{X-ray} &  & 30 unique targets \\ 
 
ROSAT & (0.2-2 keV) & 7 det, 107 UL \\ 
 
XMM-Newton & (0.2-12 keV)  & 21 det, 19 UL \\ 
 
Exosat & (2-12 keV)  & 6 det, 3 UL \\ 

Einstein & (0.2-2 keV)  & 4 det, 2 UL \\

\hline 
\textbf{Ultra-violet} & (FUV, NUV) &  \\ 
 
Galex & 1350-1780, 1770-2730 & 69 \\ 
 
 &  &  \\  
 
Swift-UVOT & \textbf{Total} & 18 \\  
Swift-UVOT & u (3072-3857) &  \\ 
Swift-UVOT &  uvw1 (2253-2946) &  \\ 
Swift-UVOT &  uvm2 (1997-2495) &  \\ 
Swift-UVOT &  uvw2 (1599-2256) &  \\ 

 &  &  \\  
 
XMM-OM & \textbf{Total} &  16\\  
XMM-OM & U (3020-3860) &  \\ 

XMM-OM & UVW1 (2495-3325) &  \\

XMM-OM & UVM2 (2070-2550)&  \\

XMM-OM & UVW2 (1870-2370) &  \\
 
 &  &  \\

HST &  \textbf{Total} & 27 \\ 
HST & Various COS 815 – 3200 

 &  \\ 
 
HST & 
STIS 1150 – 3200 in UV
 &  \\ 

 &  &  \\ 

\hline 
\end{tabular} 
\begin{tablenotes}
\item{\textbf{Notes.}} Column 2 lists the wavelength ranges covered by the various instruments and filters used for observations. Note that Swift-UVOT and XMM-OM filters do not cover exactly the same range, despite similar filter names. Column 3 lists the number of targets detected by each observatory which have flux measurements included in the catalogue. UL means targets have an upper limit value available rather than a detection.
\end{tablenotes}
\end{threeparttable}
\end{minipage}
\end{table}


\subsection{HST UV spectra}

Detailed UV spectra are available for 27 of the targets from HST pointed observations taken with the STIS and COS instruments. A subset of 13 have spectra taken with the COS G130M filter, which covers the 1100 - 1450 \AA\ wavelength range. The analysis carried out for the database was limited to G130M observations to provide the most consistent analysis, but similar spectra are available for more host stars. For example, G140L spectra for \textit{WASP-13} and \textit{WASP-18} (e.g.\ \citealt{ Fossati_2015, Fossati_2018}). The G130M spectra have been analysed in more detail to provide broadband UV flux measurements as well as flux in specific lines which are useful diagnostics of the stellar chromosphere. The flux in the lines of N \textsc{v} (1240 \AA), Si \textsc{iii} (1206 \AA), Si \textsc{iv} (1400 \AA), C \textsc{ii} (1335 \AA) were measured by multiplying the flux in each spectral bin by the wavelength width of the bin. The line flux is then the sum of all flux bins within the wave range covered by the line (see equation 1 in \citealt{France_2018}). The continuum at either side of the line was used to fit a linear continuum model and interpolate the continuum flux in the region of the line. This was then subtracted from the line flux. Line flux uncertainty was calculated by propagating the uncertainty on the flux from each spectral bin through the calculations. The N\textsc{v}, Si \textsc{iii} and Si \textsc{iv} line fluxes are unaffected by ISM absorption. However, the C \textsc{ii} line (1335 \AA) is affected. The C \textsc{ii} line fluxes in this catalogue are not adjusted to compensate for ISM absorption. It is recommended, as in \cite{France_2018}, that the other line fluxes are used when comparing activity levels of stars. Methods for estimating the line absorption are discussed in \cite{Redfield_and_Linsky_2004}.

The above procedure is not appropriate for the Lyman-alpha line at 1215.67 \AA\ because it does not account for the absorption by neutral hydrogen in the interstellar medium (ISM), which reduces the observed line flux. The flux measured directly from the spectrum is therefore unsuitable for calculating the intrinsic flux of the star or the irradiation of associated exoplanets by Lyman-alpha emission. An estimate of the intrinsic (i.e. unabsorbed by ISM) Lyman-alpha line flux was made based on the star’s observed X-ray flux using the empirical relations in \cite{Linsky_2013}. These relations are based on a sample of nearby stars where both X-ray and Lyman-alpha flux have been observed, and the unabsorbed Lyman-alpha line has been reconstructed by modelling the ISM absorption \citep{Wood_2005} or using an iterative technique \citep{France_2012}. The equations relating X-ray flux and unabsorbed Lyman-alpha flux are given in Table 2 of \cite{Linsky_2013} along with the A, B coefficients appropriate for F - G, K and M spectral types. For these Lyman-alpha unabsorbed flux estimates, the X-ray flux from XMM-Newton spectra was used. The absorbed Lyman-alpha line flux measurements are also included in the database for the purposes of assessing target feasibility for future UV observations. These are particularly relevant for studies designed to detect atmospheric evaporation using transit observations of the Lyman-alpha line. 


\subsection{Planet data}
All the planet data included in our catalogue have been retrieved from The Extrasolar Planets Encyclopedia\footnote{http://exoplanet.eu/}. The exoplenets.eu database also includes some stellar parameters which we have incorporated into our database for completeness.

\section{Exoplanets-A Database and tools} \label{sec_database}
The catalogued data and the derived quantities  discussed in Section 2 have been incorporated into a database, with an on-line user interface$^6$, as discussed in this Section (see also Appendix  \ref{app_dbexamples}). The database is currently frozen. However, we are working towards a future upgrade that would include Gaia DR4 together with the new JWST observations and links to other spectrum repositories.
\subsection{Database description}

As one of the main milestones of the Exoplanets-A project, we have implemented an archive with all the relevant information about the 205 planets and their host stars from online archives and publications. This is a Virtual Observatory compliant archive built using a special version of the SVOCat\footnote{http://svocats.cab.inta-csic.es/SVOCat/SVOCat-doc-2.2-excat2/} publishing tool.

All the data are stored in a MySQL database and can be accessed both from its webpage\footnote{http://svo2.cab.inta-csic.es/vocats/exostars/} or using Virtual Observatory protocols such as ConeSearch or SAMP. ConeSearch permits the user to restrict the search to objects in a cone centered around one position in the sky.\footnote{Example: (svo2.cab.inta-csic.es/vocats/exostars/cs.php?\\RA=172.33\&DEC=-1.45\&SR=1)} SAMP allows the database to be used interactively with other VO applications (such as TOPCAT or Aladin) and exchange results with each other in a manner seamless to the user (see Figure~\ref{fig:link} of the appendix).

The database displays one row per exoplanet (205 rows) and 700+ data columns. This number includes stellar properties, error ranges, and references to the originating material (See Section 2). The database is divided into sections to make it easier to find specific types of data and, while there are some default settings for which columns (i.e. stellar properties) are displayed, one or more user-selected sections can easily be displayed.
Broadly, the database consists of basic identification data in the sections “Names and Co-ordinates” and “Position distance Gaia DR2”. The Gaia section includes other useful stellar data such as $\mathrm{T_{eff}}$, colour and G magnitude. 
Planet data (mass, radius, $\mathrm{P_{orb}}$, eccentricity etc.) are in the sections "Exoplanets.eu".
X-ray data are in several sections. The most important is “X-ray Exoplanets-A” which includes model fits and flux measurements from XMM pointed observations. Archival data has been gathered from 3XMM, which relies on automated fitting, and the Upper Limit server for ROSAT, Einstein and Exosat data. These are in sections “X-ray 3XMM”, “X-ray ROSAT” and ”X-ray literature values” respectively.
UV photometry is in sections “UV Galex” and “UV mag Swift XMM-OM”. HST UV spectra, including line-flux measurements, are in the “UV HST Exoplanets-A” section.
For convenience when looking at the whole sample, the best available UV and X-ray data have been collected into single columns in the sections “UV preferred values” and “X-ray preferred values”. We list all the sections available in the DB, together with their provenance in Table~\ref{DBtables}. 

The database has a web-based user interface allowing simple filtering and interrogation of the contents, and the option to download as CSV-, VOtable- or JSON-format files. In addition, thanks to hyperlinks within the database, it is possible to access all the information for any planet studied by the Exoplanets-A project in an individual page. These pages contain the summarized properties of the planetary system, the exoplanet, and its host star (See Appendix A for an example). 
In particular, we include light-curves from TESS ($\sim$100 light curves have been found) and pipeline products, as well as a link to the Haute-Provence Observatory (OHP) repositories for spectra corresponding to the host star and transmission spectroscopy for the planet from ELODIE and SOPHIE can be found. We selected OHP repositories in a ‘proof of concept’ manner with the intention of including at least the ESO archives in a future upgrade. Tools have been incorporated for easy visualization. Furthermore, reductions of the spectra using Exoplanet-A's software, CASCADe\footnote{CASCADe (Calibration of trAnsit Spectroscopy using CAusal Data) is a python code used to calibrate the spectroscopic data for transiting exoplanets and to extract the transit or emission spectrum of the exoplanet. https://www.explore-exoplanets.eu/resource/cascade/}, can now be visualized and downloaded. Finally, the most important publications related to the data are listed and summarized at the end of the page.

\begin{table}
\caption{Information provenance in the DB}             
\label{DBtables}      
\centering          
\begin{tabular}{l l}     
\hline\hline       
Section Name &  Source\\
\hline                    
   Objects\tablefootmark{a} & --\\  
   Names and Coordinates &  SIMBAD     \\
   Exoplanet.eu & Encyclopaedia of exoplanetary\\
   &systems (exoplanets.eu)   \\
   Exeter Library &   Exoplanet transmission\\
   & spectroscopy from \\
   &Exeter University  \\
   Sweet-Cat &  Sweet-Cat catalogue   \\
   Position Distance Gaia DR2 &  Gaia DR2    \\
   Optical IR mag SIMBAD &  SIMBAD    \\
   Positions adjusted for PM &  Derived by us    \\
   Available Data Summary &  Derived by us    \\
   X-ray Data Available  &  XMM archive    \\
   X-ray Preferred Values &  Derived by us    \\
   X-ray Literature Values  & Literature    \\
   X-ray ExoplanetsA &  Derived by us    \\
   X-ray 3XMM &    XMM archive  \\
   X-ray ROSAT &    ROSAT archive  \\
   UV Galex &    Galex (MAST archive)  \\
   UV HST ExoplanetsA &    Derived by us  \\
   UV HST Literature &   Literature   \\
   UV mag Swift XMM OM  &  MAST/XMM archive   \\
   UV Preferred Values  &   Derived by us  \\
   Derived Quantities   &   Derived by us    \\
   Validation Flags   &  Derived by us     \\
   
\hline                  
\end{tabular}
\tablefoot{
\tablefoottext{a}{This table is a single column with the object name. This is the unique ID by which a planet is recognized and crossmatched through the whole DB.}}
\end{table}
%


\subsection{Validation of the database contents}

We have carried out many general checks to ensure the validity of the data in various ways. We have specially investigated the consistency of values when multiple sources of data are available and compared our values with those found in other catalogues. For example, all XUV measurements were plotted to check that they fall within expected ranges for stellar coronal/chromospheric emission. Flux values from our own analysis were compared to values from the literature and from automatic extractions such as the 3XMM database to check that they are consistent. Twelve of the stars in this catalogue also appear in the X-ray flux catalogue of \citet{Spinelli_2023}. The flux values from both independent analyses are in agreement within 1 $\sigma$, except for \textit{Wasp-43} and \textit{GJ 4370} which are within 2 $\sigma$. General checks were also carried out by comparing related quantities, such as spectral type and stellar effective temperature or planet orbital period and the flux received at the planet, searching for outliers and verifying that the data and derived values are correlated as expected. 


\subsection{Statistics}

 The 114 systems in our sample were selected because they had HST and/or Spitzer transit observations. This results in a sample of 205 planets, mostly with shorter orbital periods (as can be inferred from Fig~\ref{fig:Planet_Mass_vs_Semi_major_axis}), as close-in planets are more likely to be detected with the transit method. These are also the targets most likely to be suitable for transit spectroscopy and atmosphere studies with JWST. In fact, 31 of the planetary  systems in the  exoplanets-A DB are being observed with  JWST.

\begin{figure}
\includegraphics[width=88mm]{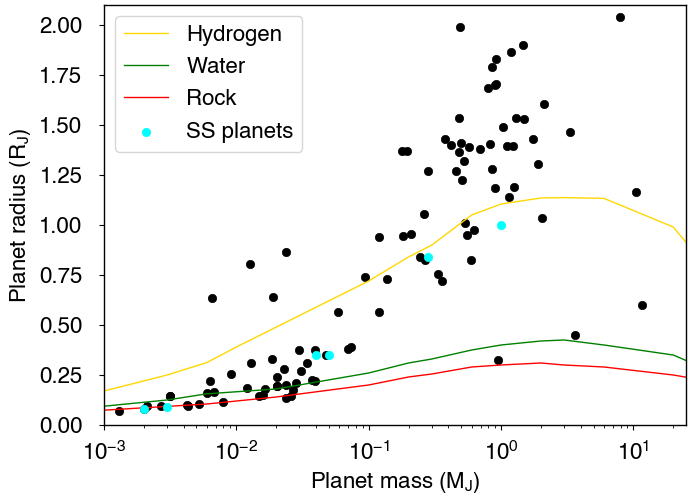}
  \caption{Radii and masses of our sample of exoplanets (black circles). Also shown are models for planets of different compositions derived by \cite{Seager07}. Solar-system planets are marked in cyan.}
  \label{fig:Planet_Mass_Jup_vs_radii}
\end{figure}

\begin{figure}
\includegraphics[width=88mm]{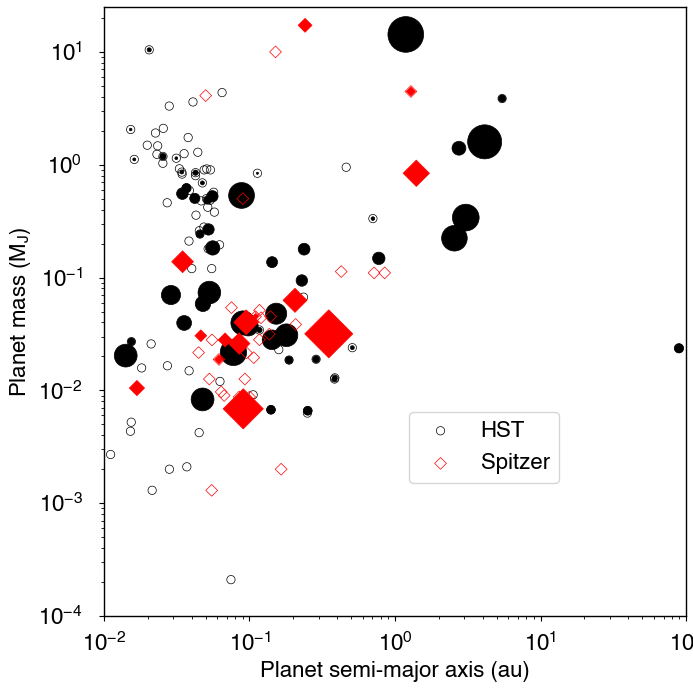}
  \caption{Masses and semi-major axes of our transiting exoplanets. Planets that have Spitzer or HST transit observations are shown in red or black respectively. Filled symbols are for planets with a value of orbital eccentricity in our database and symbol size scales linearly with eccentricity ranging from $\sim$0 to 0.7.}
  \label{fig:Planet_Mass_vs_Semi_major_axis}
\end{figure}

For each star, photometry from the SWEET-CaT and Gaia catalogue were used to fit the spectral energy distributions (using exoVOSA\footnote{\label{note10} http://svo2.cab.inta-csic.es/theory/exovosa/}). There was no IR excess in any object. The Gaia G magnitudes of our host stars range from 5.2 to 15.8 mag and the range of masses, radii and semi-major axis of the planets can be seen in Fig.~\ref{fig:Planet_Mass_Jup_vs_radii}  and Fig.~\ref{fig:Planet_Mass_vs_Semi_major_axis}. Figure~\ref{fig:Planet_Mass_Jup_vs_radii} shows about 50 rocky planets in the lower left corner according to the models for different compositions derived by \citet{Seager07}. Figure~\ref{fig:Planet_Mass_vs_Semi_major_axis} shows a clump in the top left part consisting of circularized Hot Jupiters. The limit of current detection capabilities is clearly shown by the lack of planets in the lower right part of the panel.


\begin{figure}
\includegraphics[width=88mm, height=80mm]{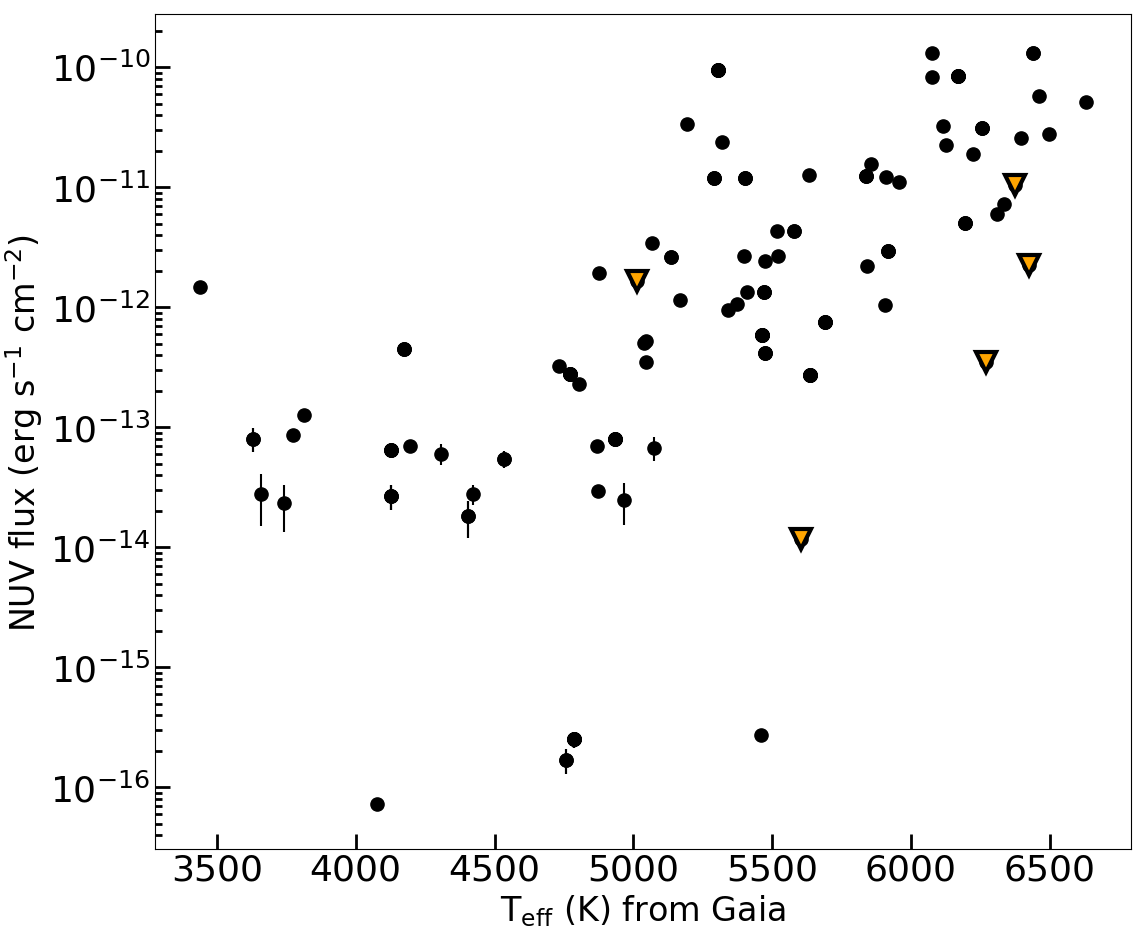}
  \caption{The distribution of detected NUV flux from the Galex (black circles) and Swift-UVOT instruments (orange triangles).}
  \label{fig:NUV_flux_vs_Teff}
\end{figure}

\begin{figure}
\includegraphics[width=88mm, height=80mm]{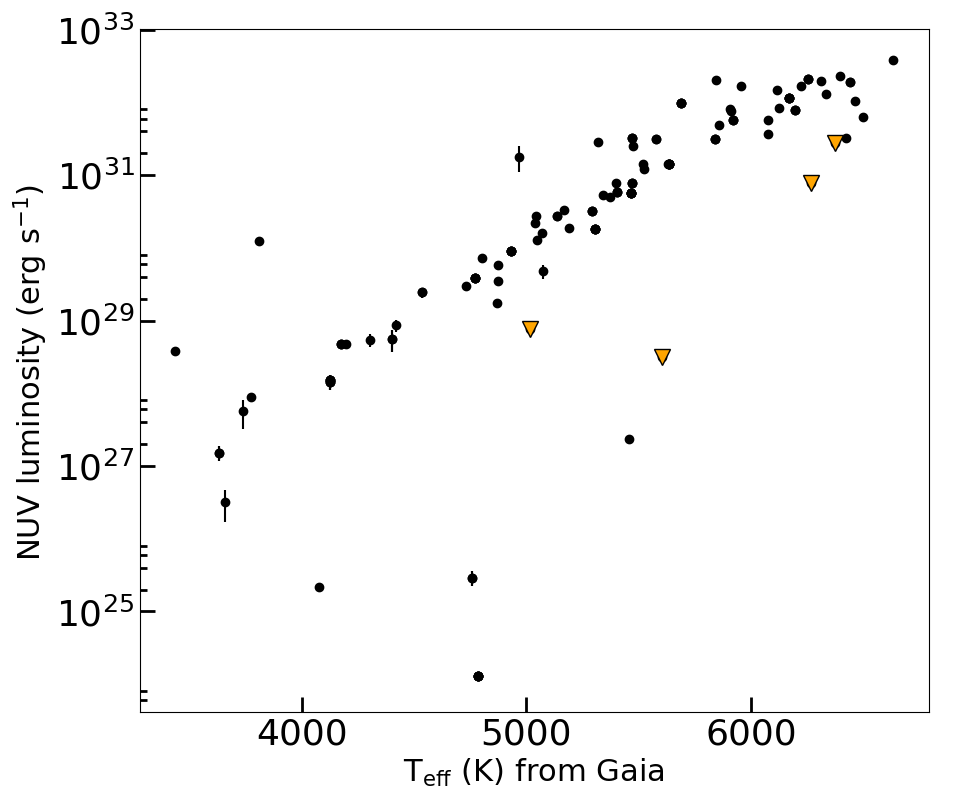}
  \caption{The NUV luminosity distribution derived from the  Galex (black circles) and Swift-UVOT observations (orange triangles).}
  \label{fig:NUV_luminosity_vs_Teff}
\end{figure}

\begin{figure}
\includegraphics[width=88mm, height=80mm]{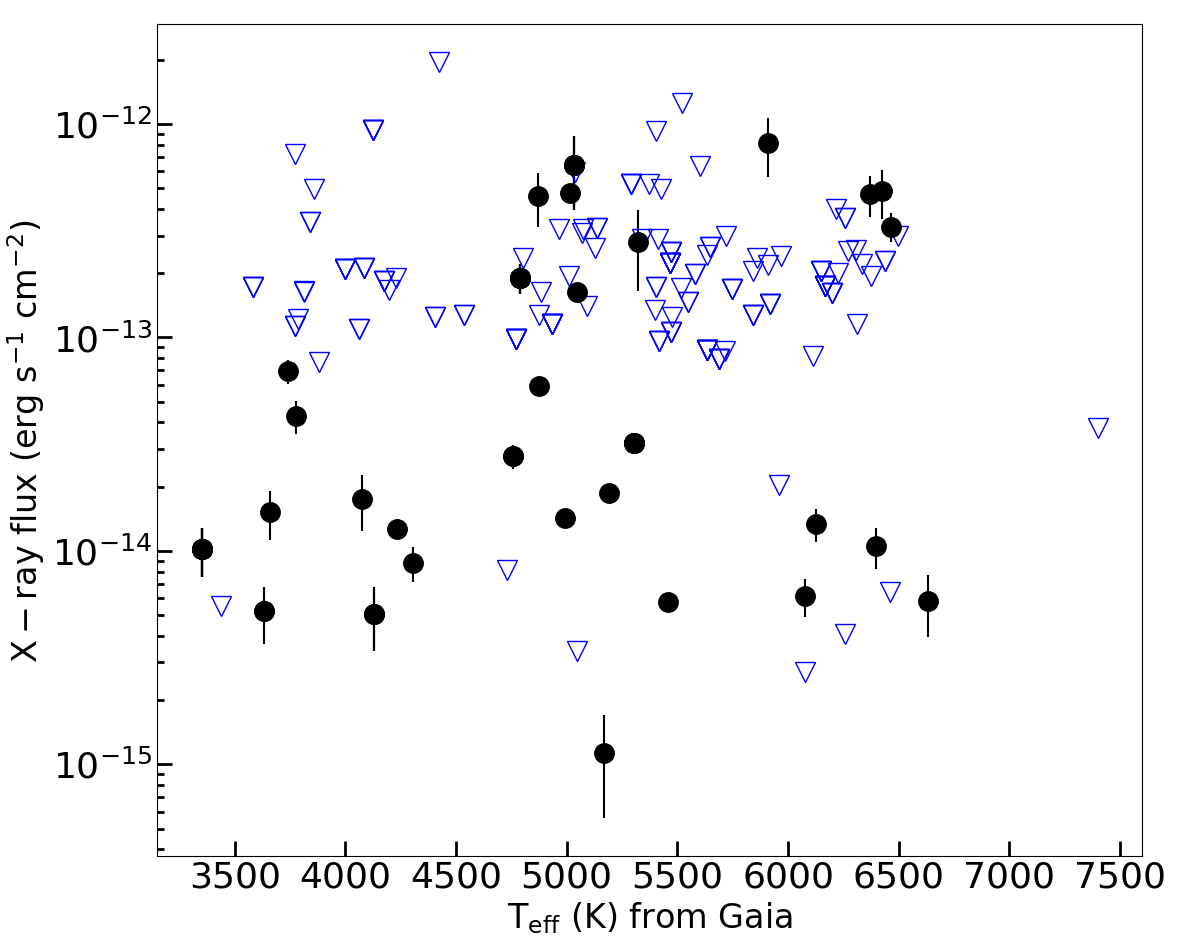}
  \caption{The distribution of detected X-ray flux as a function of stellar effective temperature shows a wide variation in flux across nearly 5 orders of magnitude. Detections are from XMM, SWIFT, ROSAT or Einstein (black circles). Also marked are upper limits (blue triangles). }
  \label{fig:Xray_flux_vs_Teff}
\end{figure}

\begin{figure}
\includegraphics[width=88mm, height=80mm]{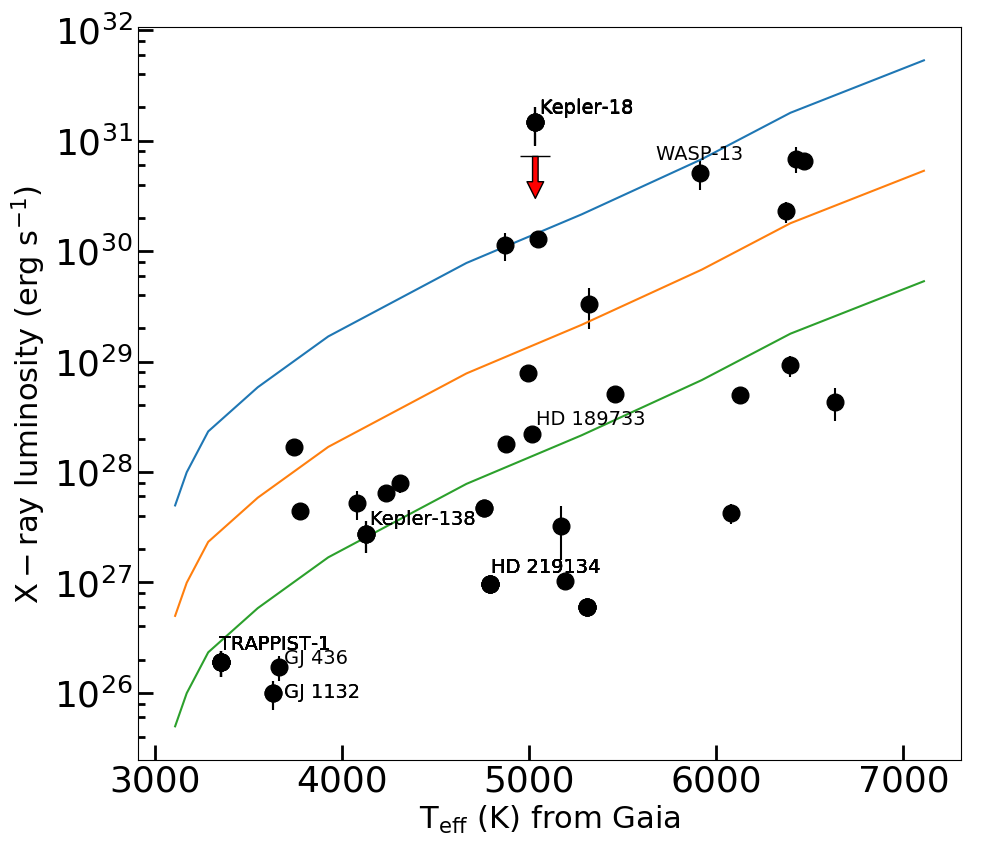}
  \caption{The distribution of detected X-ray luminosity as a function of stellar effective temperature. The blue line shows the log L$_{\mathrm{X}}$/L$_{\mathrm{bol}}$ = -3 saturation limit. The orange line is -4 and green is -5. The \textit{Kepler}-18 data point is an \textit{Einstein} 1ks exposure-time  detection from 1979. Later XMM-slew surveys with much lower exposure times found an upper limit indicated by the red arrow.    }
  \label{fig:XUV_luminosity_vs_Teff}
\end{figure}

Figure \ref{fig:XUV_det_histogram} summarises the number of stars with XUV data available, according to spectral type. A total of 30/77 stars have X-ray/NUV flux measurements in the database. These stars have a total of 52/143 known planets. Figures~\ref{fig:NUV_flux_vs_Teff} to~\ref{fig:XUV_luminosity_vs_Teff} show the range of X-ray and NUV flux and luminosity values for the whole sample, plotted according to the host-star $\mathrm{T_{eff}}$.
The detected NUV flux ranges from 1.8 $\times 10^{-14}$ up to 1.3 $\times 10^{-10}$ erg s$^{-1}$ cm$^{-2}$ (Fig.\ref{fig:NUV_flux_vs_Teff}). The intrinsic NUV luminosity calculated using the Gaia DR2 distances to the stars ranges from 3.2 $\times 10^{26}$ to 3.8 $\times 10^{32}$ erg s$^{-1}$, covering ~6 orders of magnitude (Fig.\ref{fig:NUV_luminosity_vs_Teff}). 

The X-ray detected flux ranges from 1.1 $\times 10^{-15}$ up to 1.6 $\times 10^{-13}$ erg s$^{-1}$ cm$^{-2}$ (Fig.\ref{fig:Xray_flux_vs_Teff}). The lowest flux target detected at $\sim 10^{-15}$ erg s$^{-1}$ cm$^{-2}$ is XO-2 which had an exposure time of 25 ks resulting in a very faint detection.  The X ray luminosity range is 1 $\times 10^{26}$ up to 1.5 $\times 10^{31}$ erg s$^{-1}$,  covering five orders of magnitude (Fig.\ref{fig:XUV_luminosity_vs_Teff}). The blue line shows the log L$_{\mathrm{X}}$/L$_{\mathrm{bol}}$ = -3 saturation limit \citep{Pizzolato_2003}. Stars below this line are no longer in the saturated X-ray emission phase, and the level of X-ray activity depends on the stellar rotation rate. Orange and green lines show the log L$_{\mathrm{X}}$/L$_{\mathrm{bol}}$ relation for -4 and -5 respectively. No targets more distant than 500 pc have been detected in X-ray observations (11 host stars in our sample have Gaia DR2 distances larger than 500 pc). Similarly, the NUV sample from Galex and Swift-UVOT covers a flux range of 4 orders of magnitude with very few targets detected at distances greater than 500 pc.

\subsection{ExoVOSA: fitting the observational data of exoplanets}
In addition to the database, we have developed a new tool, exoVOSA\footnote{See footnote 10}, based in VOSA (VO SED Analyzer) software \citep{Bayo08}, which aims at fitting the observational data of exoplanets. The tool is designed to perform the following tasks in an automatic manner: (i) read user photometry-tables; (ii) query several photometrical catalogs accessible through VO services (which increases the wavelength coverage of the data to be analyzed); (iii) query VO-compliant theoretical models (spectra) and calculate their synthetic photometry; (iv) perform a statistical test to determine which model reproduces best the observed data; and (v) use the best-fit model to provide bolometric luminosity (by integrating the photometry and using the known distance) and effective temperature, surface gravity and metallicity (from the model fit).
ExoVOSA is currently available for planets detected by direct images and a limited collection of 13 theoretical grids, developed for brown dwarfs and non-irradiated massive planets to fit the spectral energy distribution of exoplanets. 


\section{Discussion} \label{sec_discussion}

Here we examine what our database content may be able to tell us about the effects of the host-star high-energy (XUV) emission on the exoplanet atmosphere.
In particular, we can take advantage of the relatively large
number of exoplanetary systems for which we have compiled both XUV and planetary data, compared with many previously reported studies. This allows us to explore the relations between stellar radiation and planetary radius and atmosphere type in more detail. Also, we are directly addressing the sample for which the Exoplanets-A project has performed a uniform analysis of the atmospheres.


\subsection{The planet radius valley}

The final configuration of an exoplanet, such as terrestrial, mini-Neptune or gas giant, is largely dependent on the type of atmosphere it retains. Planet formation theory suggests that rocky planets (0.1 - 5~$\mathrm{M_{\bigoplus}}$) form with a thick H/He envelope (\citealt{Mizuno_1978, Lammer_2014, Owen_Estrada_2020}) which would greatly increase their observed radius, and render them uninhabitable to any known form of life. However, many planets in our solar system and beyond do not have thick H/He envelopes. The photoevaporation process, which has been observed in several transiting systems (\citealt{Poppenhaeger_2013, Ehrenreich_2015}), could fulfil the function of removing thick primordial H/He atmospheres from young planets, allowing a secondary thin atmosphere to form later. Stellar XUV radiation can play a role by causing heating and evaporation of atmospheres, and this process would be most effective for planets which are close-in to highly active stars \citep{Hazra_2022, Damasso_2023, Ketzer_2024}. A measured or estimated stellar $\mathrm{L_{X}}$ is needed for modelling the atmospheric mass loss rate due to XUV irradiation, and thus determining if the rate is sufficient to completely remove the H/He envelope. Stellar L$_{\mathrm{X}}$ estimates are also needed in order to study the effects the wind could have on planetary magnetospheres and atmospheres \citep{Ahuir_2020}.

Photoevaporation is not the only process that could potentially cause the radius gap. Planet formation simulations show that the distribution of material in the disk can produce a twin peaked mass distribution where planets formed close to the star are lower mass, around 3 $\mathrm{M_{\bigoplus}}$, while those beyond the ice line are rarely below 5 $\mathrm{M_{\bigoplus}}$ \cite{Venturini_2020}. The simulations showed that photoevaporation is also an important factor, and more study is needed to establish the relative importance of each process. It may be that the dominant process for determining planet radius distributions has some dependence on spectral type. \cite{Bonfanti_2024} showed that M-dwarf planet systems have a slightly different slope of radius valley compared to FGK stars, indicating that processes like gas-poor formation and inward migration of water worlds may be more important for planets around M-dwarfs. 

Atmospheric evaporation itself may also be driven by different processes. \cite{Modirrousta_Galian_2023} identified different regimes of atmospheric evaporation, where newly formed planets have atmospheric evaporation mainly driven by their high internal core temperature. XUV driven photoevaporation becomes important later, after the planet core has cooled. Boil-off driven by the stellar continuum should also be considered for low mass planets.  \cite{Affolter_2023} showed that hydrodynamic simulations are better able to reproduce the observed radius valley when the boil-off process is included. The XUV energy-limited approximation tends to underestimate atmospheric escape rates for low mass planets. It is likely that a combination of these process affects each exoplanet to varying degrees.

Here we discuss an example of how the Exoplanet-A database can be used to explore the extent to which photoevaporation could be contributing to the radius valley. Using our sample of 205 exoplanets and associated XUV data, we investigate the evidence for XUV radiation having a detectable effect on planetary H/He atmospheres. The theory of primordial atmosphere removal by photoevaporation (\citealt{Sekiya_1980a}) predicts that there will be a lack of planets with radii $\sim$2 $\mathrm{R_{\bigoplus}}$ \citep{Owen_and_Wu_2013}. Several studies have detected this gap in samples of exoplanets (e.g. \citealt{Fulton_2017, van_Eylen_2018, Owen_Estrada_2020}). We start by investigating whether our sample also shows the radius gap.

\begin{figure}
\includegraphics[width=88mm, height=88mm]{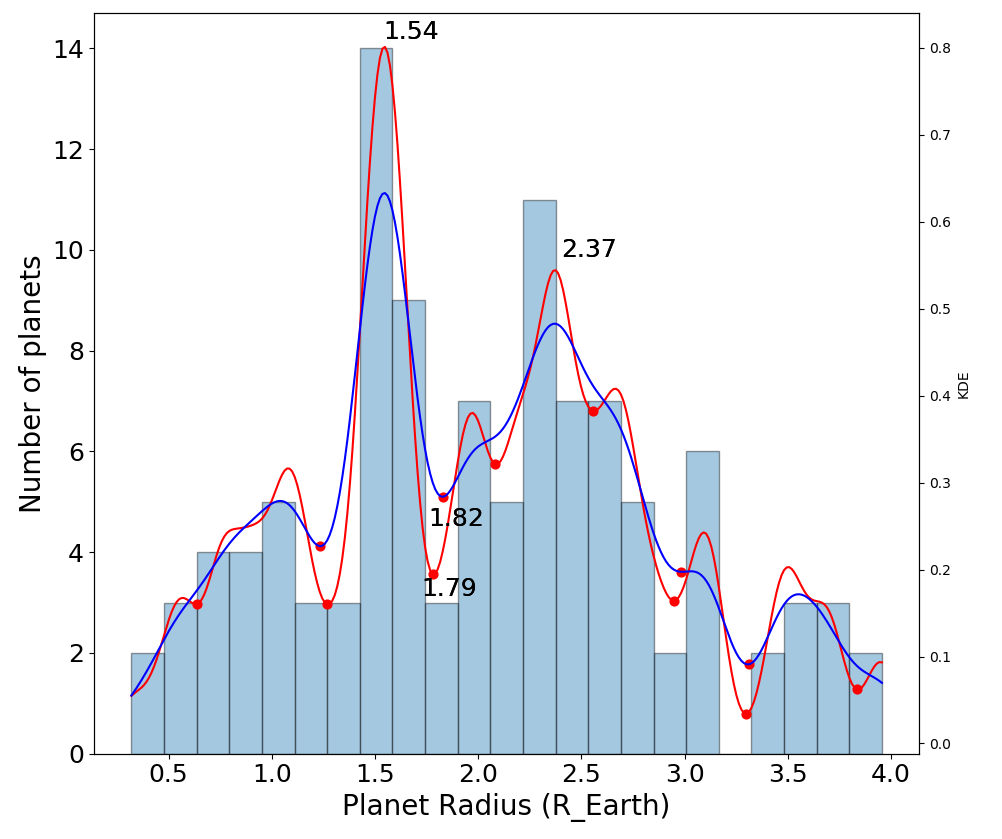}
  \caption{The planet radius histogram showing the radius gap at $\sim$1.8 Earth radii. The blue and red lines are kernel density estimations of the smoothed underlying distribution and identify the location of the radius separating the two planet populations ($\sim 1.8$), and the peaks of the two planet populations at 1.5 and 2.4 Earth radii.}
  \label{fig:Planet_radius_histogram}
\end{figure}

Figure~\ref{fig:Planet_radius_histogram} shows the distribution of planet radii in the Exoplanets-A sample. The data are binned in logarithmic bins of 0.16 $\mathrm{R_{\bigoplus}}$ to match the binning used for Fig 4 in \cite{van_Eylen_2018} for comparison.  Both samples show a twin peaked distribution, with the “radius-valley” at $\sim$2 $\mathrm{R_{\bigoplus}}$.

First, we consider whether the radius distribution is really multimodal. Although we see a clear dip in the sample at the expected radius, there are also several other dips, and it is important to decide which ones are statistically significant. The Hartigan dip test of multimodality \citep{Hartigan_1985} was applied to the log $R_{\mathrm{P}}$ data in the range $R<3\mathrm{R_{\bigoplus}}$ following the method of \cite{Fulton_2017}. Our sample returned a $p$-value of 0.042 that the distribution is unimodal, which is larger than the 0.0014 reported by \cite{Fulton_2017}, but still shows that the probability that this sample is multimodal is >95 percent. 

Next we identify the location of the valley and the peak of the 2 distinct planet populations. The binning of the histogram could cause gaps to appear due to the low number of data points in the sample. Rather than identify the location of the valley directly from the histogram, we used a Kernel density estimator (KDE) analysis \citep{Rosenblatt_1956, Parzen_1962} to find the location of the valley in this sample. This method smooths out noise in the data so that only significant peaks and dips are detected.

The minimum in the radius valley is located at 1.8 $\mathrm{R_{\bigoplus}}$.
The peak of the two distributions either side of the valley are at 1.5 $\mathrm{R_{\bigoplus}}$ (super-Earths) and 2.4 $\mathrm{R_{\bigoplus}}$ (mini-Neptunes).
The bandwidth used by the KDE method determines how sensitive the result is to noise in the data. The bandwidth in the KDE method is equivalent to the bin width for a histogram. A bandwidth that is too large could smooth out real features. To check the effect of bandwidth we used the "Cross Validation gridsearch" test implemented in the Python library \textit{sklearn} \citep{Pedregosa_2011}. This method repeatedly splits the data randomly into training and testing samples so that the validity of the solution can be tested without having to remove a portion of the training data for testing purposes. The cross-validation found the best bandwidth is 0.078 which gives a radius valley located at 1.79 $\mathrm{R_{\bigoplus}}$ shown by the red line in Fig. \ref{fig:Planet_radius_histogram}.  Using a KDE with a larger bandwidth of 0.12 shifts the identified minimum to slightly larger radius of 1.82 $\mathrm{R_{\bigoplus}}$ as shown by the blue line in Fig. \ref{fig:Planet_radius_histogram}.


Having confirmed the presence of the radius gap in our planet population, we next investigate what could have caused it. The photoevaporation model (e.g. \citealt{Owen_Estrada_2020}) predicts a radius gap at this range because the planet radius determines how strongly a planet can hold on to its H/He atmosphere. Low-mass planets with a core at or below the 1.8 $\mathrm{R_{\bigoplus}}$ limit may lose their atmosphere via a rapid hydrodynamical process if subjected to stellar XUV radiation, so after several Gyrs, all that remains is a stripped core (e.g. \citealt{Becker_2020}).  Planets with a core above the $\sim$1.8 $\mathrm{R_{\bigoplus}}$ limit have a stronger gravitational potential and are able to retain some of their envelope. They are more likely to remain with a radius larger than $\sim$1.8 $\mathrm{R_{\bigoplus}}$, since only a small mass of retained H/He atmosphere can greatly increase the observed planetary radius. There are four planets in our sample which appear to be within the radius gap, between 1.7 and 2 $\mathrm{R_{\bigoplus}}$. These are 55 Cnc e, \textit{Kepler}-9 d, \textit{Kepler}-20 b and \textit{Kepler}-11 b.



\begin{figure}
\includegraphics[width=88mm, height=88mm]{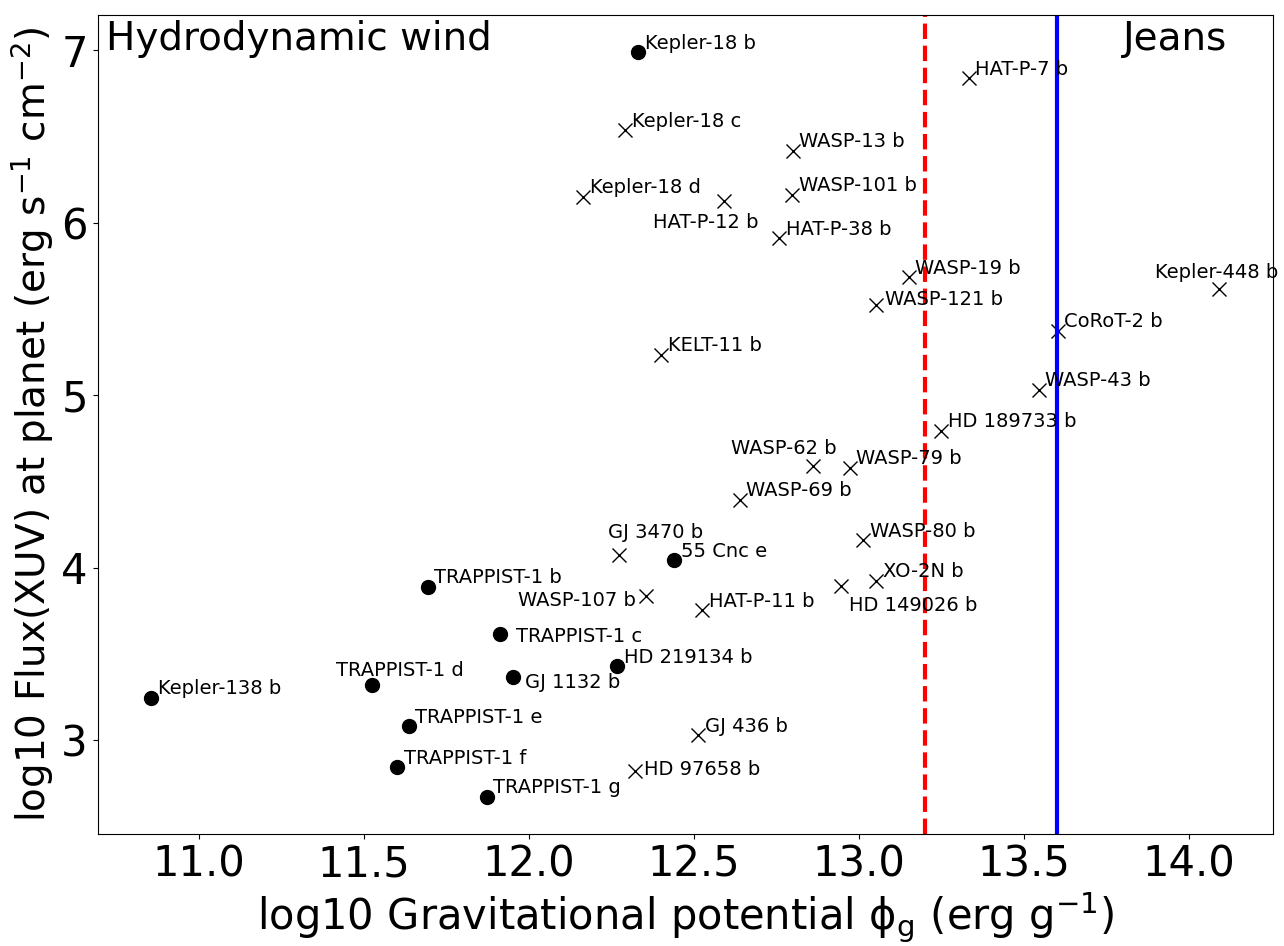}
  \caption{The XUV flux received by planets compared to their gravitational potential in order to determine the possible atmospheric escape process following the method in \cite{Salz_2016}. The rapid hydrodynamical escape process applies to the left of the red dashed line. Slower, Jeans escape applies to the right of the blue solid line. Markers are crosses for planets above the radius gap and circles for planets below the radius gap.}
  \label{fig:XUV_flux_at_planet_vs_grav_potential_atmospheric_escape}
\end{figure}

We investigate the likely evaporation processes applicable to the planets. 
The rate of evaporation, and whether or not it will be sufficient to completely remove the primordial H/He atmosphere, depends on two main factors. The first factor is the planet's gravitational potential energy, which determines the escape regime it will be in if subjected to XUV radiation. For example, rapid hydrodynamic escape, or slow Jeans escape. The second factor is the XUV flux incident at the planet which determines the rate of evaporation, given the escape regime applicable to that particular planet.  Fig.\ref{fig:XUV_flux_at_planet_vs_grav_potential_atmospheric_escape} shows the $log$ gravitational potential energy of the planets which have observed X-ray flux.

The planets gravitational potential energy $\Phi_{\mathrm{g}}$ was calculated using equation \ref{eqn:Grav_potential} and mass and radius parameters from the EU exoplanet archive.

\begin{equation}\label{eqn:Grav_potential}
   \Phi_{\mathrm{g}} = -(GM)/r
\end{equation}

The XUV flux at each planet is derived from the observed X-ray flux as follows. The observed X-ray flux was converted to luminosity based on the star-Earth distance from Gaia DR2 and using equation \ref{eqn:Flux_to_luminosity}. The EUV is not observed, but can be estimated from the X-ray luminosity using the empirical relation (equation.\ref{eqn:Lx_to_Leuv}) from \cite{Sanz_Forcada_2011}.

\begin{equation}\label{eqn:Flux_to_luminosity}
    L = 4\pi\ D^{2}f_{\mathrm{x}}
\end{equation}

\begin{equation}\label{eqn:Lx_to_Leuv}
    \log \mathrm{L_{EUV}} = 4.80 + (0.860\times \log \mathrm{L_{X}})
\end{equation}

The observed X-ray and estimated EUV luminosity are added to get the combined XUV luminosity. The flux at the planet is then calculated from the stellar XUV luminosity and the planet semi-major axis. This assumes the effect of orbital eccentricity is negligible.

There are 30 planets to the left of the red line which indicates that they have weak enough gravity so that atmospheric escape would take place via the rapid hydrodynamical loss process. To the right of the red line, only the slower Jeans escape process would be possible due to the planets stronger gravitational potential energy. The markers are plotted as circles for planets below the $\sim$1.8 $\mathrm{R_{\bigoplus}}$ radius gap. Planets with a larger radius are plotted as crosses and are planets that may still have a remaining primordial atmosphere undergoing evaporation. Those planets plotted as circles have likely had their atmospheres entirely evaporated.


Fig.\ref{fig:XUV_flux_at_planet_vs_grav_potential_atmospheric_escape} shows that, for this sample, is mostly the planets receiving less XUV flux which have low radius. One possible explanation could be that they are in older systems where the evaporation process has already taken place over a long period, and the stars X-ray luminosity has declined due to spin down (e.g. \citealt{Wright_2011, Johnstone_2020}). The majority of planets in the lower flux region belong to the \textit{Trappist-1} system, which is relatively old at 7.6$\pm$2.2 Gyr \citep{Burgasser_2017}, and is known to have an X-ray luminosity of $\mathrm{L_{X}/L_{bol}} = 2-4\times10^{-4}$ \citep{Wheatley_2017} which is lower than the typical saturation limit of $10^{-3}$. Planets receiving higher XUV flux are more likely to be orbiting active younger stars, assuming a sample with a similar spread of orbital radius and spectral type, so the evaporation process may not yet have had time to remove the whole envelope. More work is needed to establish accurate ages for host stars to be able to investigate this question. 

An important caveat to this analysis is that it does not include Roche lobe effects, which can greatly increase the mass loss rate for planets very close to their host star. For example, \textit{WASP-121 b} has a much higher rate of atmospheric loss than Fig.\ref{fig:XUV_flux_at_planet_vs_grav_potential_atmospheric_escape} would indicate \citep{Koskinen_2022, Huang_2023}.

\subsection{Evidence for the photoevaporation model from multi-planet systems}
The present day radius of a planet is a result of the cumulative mass loss that has taken place throughout it's evolution. One way to investigate whether this could be the cause of the radius valley is to estimate the total mass lost over the lifetime of the planet. For the planets where the incident X-ray flux is known, it is possible to estimate the atmospheric mass loss rate using the energy-limited escape approximation \citep{Watson_1981, Lammer_2003}. More accurate mass loss rates based on grids of hydrodynamic simulations \citep{Kubyshkina_2018_ApJ_letter,  Krenn_2021} are now available. Stellar XUV radiation can also potentially cause loss of liquid water from planets in the habitable zone of highly active stars (e.g. \citealt{Wheatley_2017}; \citealt{Johnstone_2020}). Despite many improvements to atmospheric mass loss rate estimates given the present-day conditions, a major difficulty with calculating the cumulative effect is that stellar ages remain highly uncertain for most stars. So, while the current mass-loss rate can be estimated, the star's XUV evolution, and the long-term effect of photoevaporation on the planet cannot be determined with accuracy.

\begin{figure}
\includegraphics[width=88mm, height=88mm]{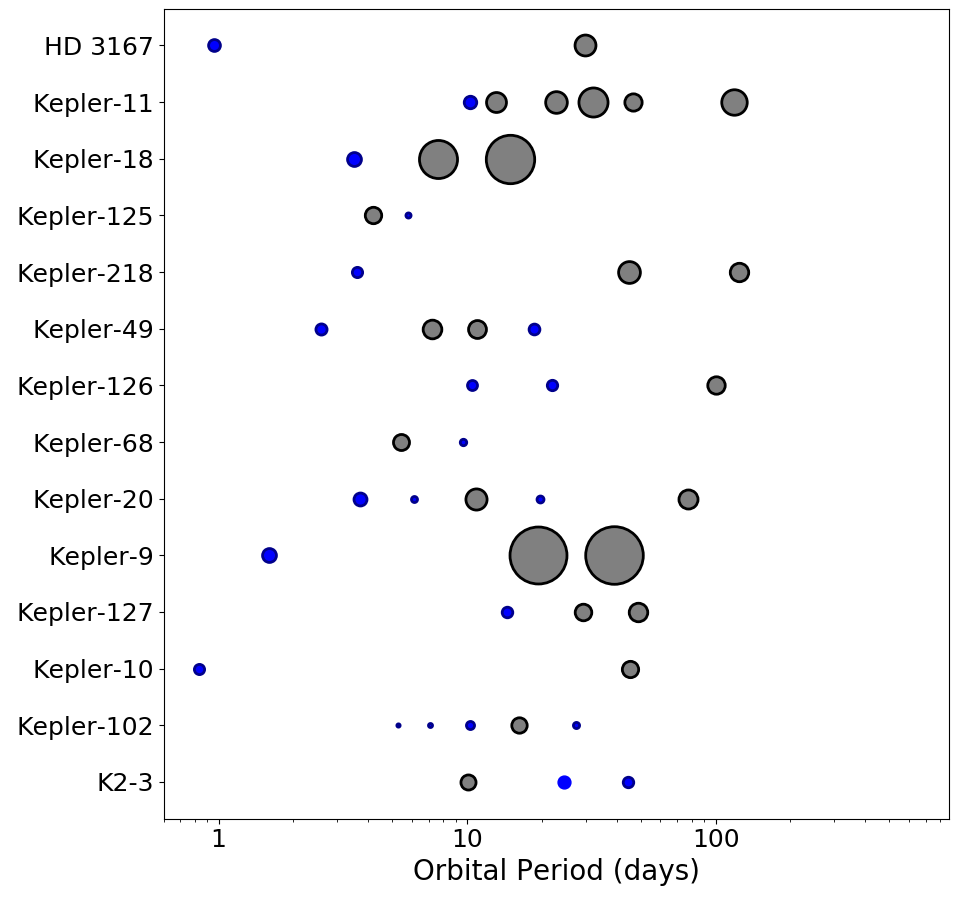}
  \caption{Architecture for the systems tested with the photoevaporation model. Planets below the 1.8 $\mathrm{R_{\bigoplus}}$ radius gap are blue and planets above the gap are grey with black outline. Marker size is scaled linearly to indicate planet radius, ranging from 0.5 to 8.3 $\mathrm{R_{\bigoplus}}$. }
  \label{fig:Systems_with_planets_above_and_below_the_radius_gap}
\end{figure}

To avoid this limitation, we investigate the comparative mass-loss of multiplanet systems in our sample. \cite{Owen_Estrada_2020} developed the method of using multiplanet systems to test the photoevaporation model. This method relies on comparing planets above and below the radius gap in the same planetary system, and has the advantage of rendering it unnecessary to know the complete XUV radiation history of the star, since both planets have been subjected to the same XUV history, once the planetary distances from the star have been taken into account.
In our sample there are 14 systems with multiple planets straddling the radius gap that could potentially be used as a test of the photoevaporation model. 8 of these systems are not in the samples tested in \cite{Owen_Estrada_2020}, and some systems have additional planets, e.g. \textit{Kepler}-102f is included in our sample. In Fig.\ref{fig:Systems_with_planets_above_and_below_the_radius_gap}, the systems which have planets above and below the radius gap are plotted as a function of the orbital period. Planets with a radius < 1.8 $\mathrm{R_{\bigoplus}}$ are plotted in blue and greater than 1.8 $\mathrm{R_{\bigoplus}}$ in grey. The marker sizes are scaled according to the planet radius.

The method considers a planetary system as a whole, and relies on assuming that the super-Earth is below the radius gap because it received just enough XUV flux to completely evaporate its atmosphere. The mass-loss timescale for the mini-Neptune must be longer than this, since it did not lose all of its envelope. After the difference in received XUV flux is taken into account, by comparing the planets' orbital periods (and hence star-planet distance), the only remaining factor which would increase the mass-loss timescale of the mini-Neptune is its mass, and hence its stronger gravity which could retain its atmosphere. Equation \ref{eqn:Minimum_mass} \citep{Owen_Estrada_2020} gives the minimum mass required for the mini-Neptune (M$_{g}$) to have retained its primordial atmosphere. If the measured planet mass is greater than this, the system is compatible with the photoevaporation model. 

\begin{equation}\label{eqn:Minimum_mass}
    M_{g} \geq 5.1 M_{\bigoplus} \Bigg(\frac{R_\mathrm{{R}}}{1.5R_{\bigoplus}}\Bigg)^{4}\Bigg(\frac{a_{R}}{a_{g}}\Bigg)
\end{equation}

In contrast, if the planet above the radius gap has a lower mass than the minimum for atmosphere retention, then it must have retained its primordial atmosphere because of some factor that is not accounted for in the photoevaporation model, such as orbit migration or planetary magnetosphere, or the super-Earth lost its atmosphere due to some other process than photoevaporation. In Equation \ref{eqn:Minimum_mass}, $R_\mathrm{{R}}$ is the radius of the rocky planet, and $a_{R}$ and $a_{g}$ are the orbital distance of the rocky and gas (mini-Neptune) planet respectively.


\begin{table*}
\begin{minipage}{170mm}
	\centering
	\caption{Table of minimum mass values for planets above the radius gap.}
	\label{tab:Owen_test_min_planet_mass_photoevaporation}

\begin{threeparttable}
\begin{tabular*}{\textwidth}{cccc|cccc}

\hline 
\textbf{Planet above} & \textbf{Radius} & \textbf{P orb} & \textbf{Measured mass} & \multicolumn{4}{c} {\textbf{Minimum mass prediction from each planet below 2 $\mathrm{R_{\bigoplus}}$}}  \\

\textbf{radius gap} & ($\mathrm{R_{\bigoplus}}$) & (days) & ($\mathrm{M_{\bigoplus}}$) & ($\mathrm{M_{\bigoplus}}$) &&&\\
\hline 
&&&& \textit{Kepler}-102 b & \textit{Kepler}-102 c & \textit{Kepler}-102 d & \textit{Kepler}-102 f \\

\textit{Kepler}-102 e & 2.22 & 16.15 & 8.93 & LM & 0.12 & 2.73 & No Solution \\ 
\hline

 &  &  &  & \textit{Kepler}-68 b & \textit{Kepler}-68 d &  &  \\ 
 
\textit{Kepler}-68 c  & 0.93 & 9.6 & 5.97 & 3.75 & Radius unknown &  \\

\hline

 &  &  &  & \textit{Kepler}-49 d & \textit{Kepler}-49 e &  &  \\ 

\textit{Kepler}-49 b & 2.69 & 7.2 & - & 1.0 & 1.0 &  &  \\ 

\textit{Kepler}-49 c & 2.58 & 10.91 & - & LM & LM &  &  \\

\hline 
 &  &  &  & \textit{Kepler}-125 c &  &  &  \\ 

\textit{Kepler}-125 b & 2.37 & 4.16 & - & 1.0 &  &  &  \\

\hline 
 &  &  &  & \textit{Kepler}-11 b &  &  &  \\

\textit{Kepler}-11 c & 2.87 & 13.03 & 2.9 & No Solution &  &  &  \\ 
\textit{Kepler}-11 d & 3.12 & 22.69 & 7.3 & 4.51 &  &  &  \\ 
\textit{Kepler}-11 e  & 4.19 & 31.99 & 9.53 & 2.02 &  &  &  \\ 
\textit{Kepler}-11 f   & 2.49 & 46.69 & 1.99 & 1.59 &  &  &  \\ 
\textit{Kepler}-11 g    & 3.67 & 118.38 & 301.9 & 0.33  &  &  &  \\

 \hline 
 &  &  &  & HD 3167 b &  &  &  \\ 

HD 3167 c & 3.02 & 29.85 & 9.78 & 0.13 &  &  &  \\

 \hline 
 &  &  &  & \textit{Kepler}-10 b &  &  &  \\ 

\textit{Kepler}-10 c & 2.35 & 45.29 & 7.36 & LM &  &  &  \\

 \hline 
 &  &  &  & \textit{Kepler}-20 b & \textit{Kepler}-20 e & \textit{Kepler}-20 f &  \\ 

\textit{Kepler}-20 c & 3.05 & 10.85 & 12.75 & 3.49 & 0.35 & 2.839 &  \\ 

\textit{Kepler}-20 d & 2.74 & 77.61 & 10.06 & 0.23 & LM & 0.23 &  \\ 
 
 \hline 
 &  &  &  & K2-3 c & K2-3 d &  &  \\ 

K2-3 b & 2.17 & 10.05 & 6.47 & No Solution & No Solution &  &  \\

  \hline 
 &  &  &  & \textit{Kepler}-9 d &  &  &  \\ 

\textit{Kepler}-9 b & 8.20 & 19.22 & 43.5 & 0.997 &  &  &  \\ 
\textit{Kepler}-9 c & 8.29 & 38.97 & 29.9 & 0.769 &  &  &  \\

  \hline 
 &  &  &  & \textit{Kepler}-18 b &  &  &  \\ 

\textit{Kepler}-18 c & 5.48 & 7.6 & 17.20 & 5.36 &  &  &  \\ 
\textit{Kepler}-18 d & 6.97 & 14.86 & 16.29 & 2.14 &  &  &  \\
 
  \hline

 &  &  &  & \textit{Kepler}-126 b & \textit{Kepler}-126 c &  &  \\ 
\textit{Kepler}-126 d & 2.51 & 100.28 & - & 0.3$^{*}$ & 0.96$^{*}$ &  &  \\

 \hline 
 &  &  &  & \textit{Kepler}-127 b & \textit{Kepler}-127 d &  &  \\ 
\textit{Kepler}-127 c & 2.39 & 29.39 & - & - & 1.79$^{*}$ &  &  \\ 
\hline

 &  &  &  & \textit{Kepler}-218 b &  &  &  \\ 

\textit{Kepler}-218 c & 3.13 & 44.70 & - & 1.0$^{*}$ &  &  &  \\ 
\textit{Kepler}-218 d & 2.66 & 124.5 & - & LM &  &  &  \\
  \hline

\end{tabular*} 
\begin{tablenotes}
\item{\textbf{Notes.}}
The planets listed in the left half of the table are above the radius gap and have an observed mass listed in column 4. These can be compared to their minimum mass predicted from the planets below the period gap (right hand section of the table). Systems marked "No Solution" are not compatible with the photoevaporation model. Note that the mass values in the right half of the table are predictions for the mass of planets in the left half, not masses of the planets in the right half. Minimum mass values with an asterisk had estimated solar mass values used in the calculation. Planets which do not constrain the minimum mass are marked LM. The minimum masses are estimated using the \cite{Owen_Estrada_2020} code.
\end{tablenotes}
\end{threeparttable}
\end{minipage}
\end{table*}


The 14 planetary systems were input to the code publicly available at \textit{https://github.com/jo276/EvapMass}. The input includes stellar and planetary properties (radius, orbital period, etc.). The output from the program is the minimum required mass of the planet above the radius-gap. Table \ref{tab:Owen_test_min_planet_mass_photoevaporation} gives the results and compares the measured and required masses of each planet. Five systems do not yet have planet mass measurements available to compare to the predicted minimum mass, and thus they could not be tested in this way.

A further complication for \textit{Kepler}-126, 127 and 218, is that there are no stellar mass data available in the Exoplanets-EU catalogue. For these 3 systems the stellar mass was estimated from the stellar radius and $\mathrm{T_{eff}}$ using the stellar main-sequence relations in \cite{Eker_2018}. These minimum mass results in the table are marked with an asterisk.


Of the 9 remaining systems we were able to test, we find that 5 of the systems are completely compatible with the photoevaporation model. A further 2 systems, both with more than two planets, had most of the planets compatible with photoevaporation, but with the following exceptions. \textit{Kepler}-102 f is exterior to “e” but is below the radius gap. No core mass (for "e") could be found that would enable “e” to have a longer mass-loss timescale than "f". However, a small increase in $\mathrm{P_{orb}}$ for planet “e” from 16.15 to 19 days would produce a compatible result. This indicates that \textit{Kepler}-102 e is only marginally incompatible with photoevaporation. 

The \textit{Kepler}-11 system has 5 known planets above the radius gap. Four of the planets have predicted minimum masses less than the measured mass except for planet "c", so they are compatible with the photoevaporation theory. The exception is planet "c". This may indicate that some process has slowed mass loss for planet "c" which has not operated on planet "b", since they are at similar orbital periods (b = 10 days and c = 13 days).

Systems where the larger radius planet does not have a longer mass-loss timescale than the planet below the radius gap are marked with as "No Solution” in Table \ref{tab:Owen_test_min_planet_mass_photoevaporation}. These systems are incompatible with the photoevaporation model.

The K2-3 system seems to be the clearest example of a contradiction to the photoevaporation model. The inner planet “b” is above the radius gap. The 2 exterior planets “c” and “d” are both below the radius gap (see bottom row of Fig. \ref{fig:Systems_with_planets_above_and_below_the_radius_gap}). Planet b does not have enough mass to have retained it's primordial atmosphere, given that it was subjected to the same XUV irradiation history as planet c and d. This could indicate that some factor not included in the model has increased the mass loss timescale of planet “b”, such as inward migration, or a strong magnetosphere which has protected the atmosphere. A similar analysis by \cite{Diamond_Lowe_2022} reached the same conclusion.

There were 5 planet combinations which did not place any constraint on the minimum mass. These are recorded as ‘LM’ in Table \ref{tab:Owen_test_min_planet_mass_photoevaporation}. "LM" (lower mass) refers to cases where no lower mass bound for the gas planet's core could be found in the range $>0.1 \mathrm{M_{\bigoplus}}$ (See Fig 1 \cite{Owen_Estrada_2020}).  This means that even a planet with a mass $<0.1 \mathrm{M_{\bigoplus}}$ would have a longer mass-loss timescale than the planet below the radius gap. The low radius planet would be easily stripped of its envelope, and the strength of photoevaporation required to do that would have no significant effect on the planet above the radius gap, regardless of its mass.

In general, we find that most systems in our sample are compatible with the photoevaporation model, but K2-3 could be a notable exception.




\section{Conclusions} \label{sec_conclusions}

--We have compiled a catalogue of host-star characteristics that includes basic properties for all 114 stars associated with the exoplanets in the ExoplANETS-A sample. It contains information on the optical and IR bandpasses for almost all the stars plus X-ray detections for $\sim 25 \%$, UV photometry for $\sim 60 \%$, and UV spectra for $\sim 25 \%$ of the stars.

--With this catalogue we have implemented a database which includes X-ray and OHP spectra, TESS light curves and other useful information on top of the catalogue itself. We have also implemented the tool exoVOSA which is able to fit the spectral energy distribution of exoplanets using a  collection of theoretical grids, developed for brown dwarfs and non-irradiated massive planets

--We have used this database information to study the effects of the host-star high-energy (XUV) emission on the exoplanet atmosphere. In particular, we have studied the planet radius valley which we find  is located at 1.8$\mathrm{R_{\bigoplus}}$, in agreement with previous studies. In addition, we have used the multiplanet systems in our sample to test the photoevaporation model. Only one system, K2-3, out of 14 systems poses a contradiction to the photoevaporation model.
The fact that the inner planet of the system is above the radius gap and the  two exterior planets are both below the radius gap may be indicating that an inward migration happened or that the planet has a strong magnetosphere which has protected the atmosphere. In any case, it  would  seem  that some factor or a mixture of factors not included in the photoevaporation model has increased the mass-loss timescale of the inner planet.

The host-star information included in our database complements the uniform set of exoplanet HST and Spitzer visible/IR spectra and associated  retrievals of atmospheric properties produced by the ExoplANETS-A project. The exoplanet and stellar resources compiled and generated by ExoplANETS-A form a sound basis for current JWST observations and future work in the era of Ariel.

\begin{acknowledgements}
We thank the reviewer for many helpful comments which improved the manuscript.
  The research leading to these results has received funding from the European Union's Horizon 2020 Research and Innovation Programme, under Grant Agreement nº 776403.
This research has been funded by the Spanish State Research Agency (AEI) Projects No.ESP2017-87676-C5-1-R and No. MDM-2017-0737 Unidad de Excelencia “María de Maeztu”- Centro de Astrobiología (INTA-CSIC).
ALL DBS  USED.

\end{acknowledgements}

%
%

%

\begin{appendix}

\section{Examples of database usage: Visualizing¬¬ the most important information for a planet/host-star} \label{app_dbexamples}

\begin{figure*}
\includegraphics[width=\hsize]{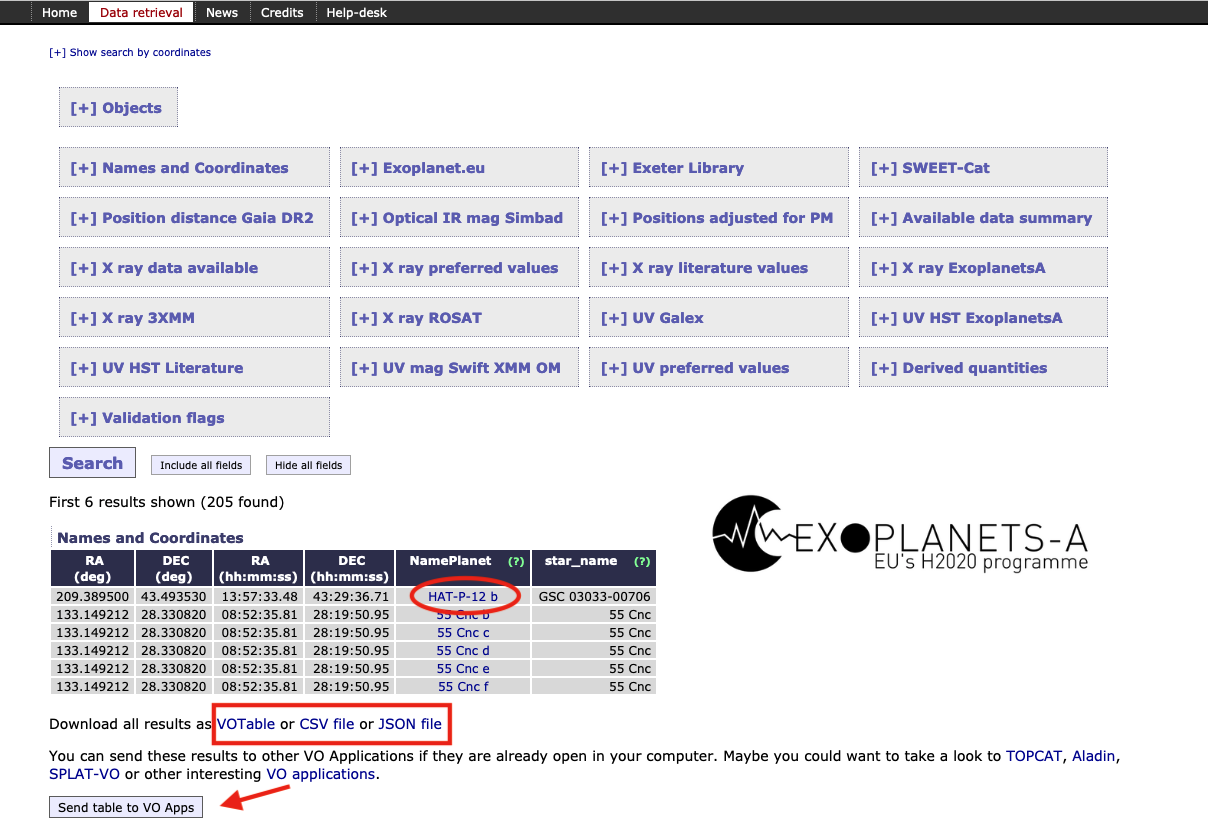}
  \caption{Default look of the database webpage. On the upper part grey rectangles with plus symbols are the different sections from where the specific fields can be marked to be displayed. Marked in red are the link to individual pages for a planet, the download available formats and the button to send the data to other VO applications.}
  \label{fig:link}
\end{figure*}

   \begin{figure*}
   \centering
   \includegraphics[width=\hsize]{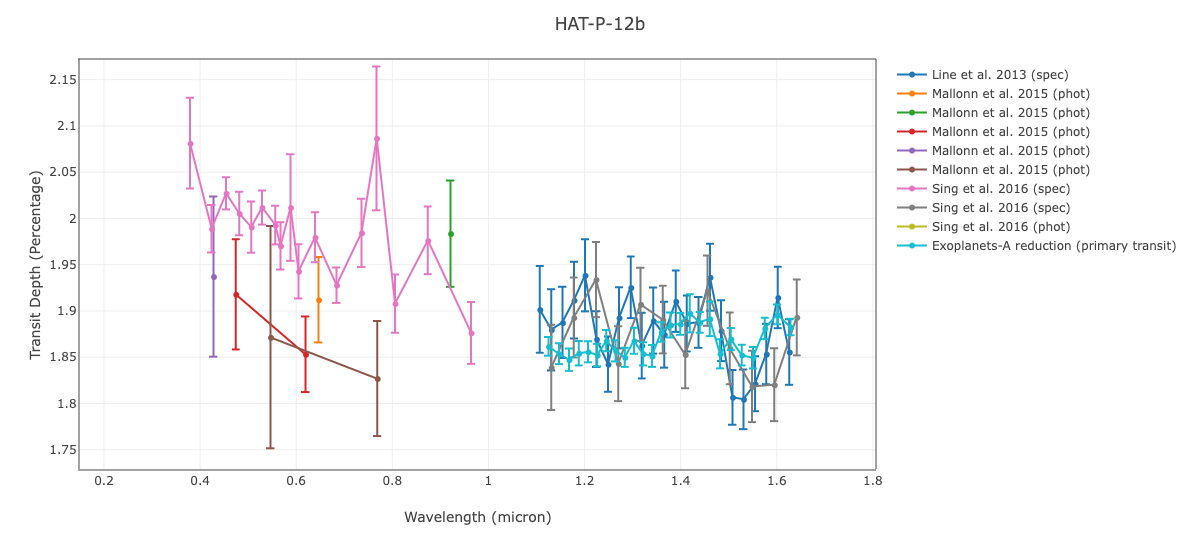}
      \caption{An example of the spectra visualization for planet HAT-P-12b with the Exoplanets-A reduction shown in cyan.
             }
         \label{fig:spectra}
   \end{figure*}

A summarized version of the planet, the whole system, and the host-star information can be accessed through individual pages by means of an hyperlink on the planet name (see Fig.\ref{fig:link}). There is an individual page per planet\footnote{An example can be seen here: http://svo2.cab.inta-csic.es/vocats/exostars/fichas/?id=222}.

In the upper part of the page there is a summary of the star properties and main magnitudes, and the system planets parameters. Then comes some external data for the host star. These data include the TESS light curves, and direct access to some TESS pipeline files for direct download, and several links that lead to predefined queries in the OHP archives for the available ELODIE and SOPHIE spectra and CCF catalogues.
At the bottom of the individual webpage  we find the information on the planet. First there is a selection of the planet parameters from the database and then an interactive visualization of the exoplanet transit depth records from NASA Exoplanet Archive (2850 records are available). Our own CASCADe reduction is also available here to visualize and to download (see Figure~\ref{fig:spectra}).
Finally there is a table summarizing and with direct access to reference papers, facility and instrument of the observations.

\end{appendix}

\end{document}